\documentclass{article}

\usepackage[final]{main}
\usepackage{natbib}
\usepackage{listings}
\usepackage[utf8]{inputenc} 
\usepackage[T1]{fontenc}    
\usepackage{booktabs}       
\usepackage{microtype}      
\usepackage{xcolor}         
\usepackage{wrapfig}
\usepackage{graphicx}
\usepackage{esvect}
\usepackage{mathrsfs}
\usepackage{fancyvrb}
\usepackage{fvextra}
\usepackage[hyphens]{url}
\usepackage{hyperref}
\usepackage[hyphenbreaks]{breakurl}

\title{Generative AI Policy and Governance Considerations for Health Security in Southeast Asia}

\makeatletter
\newcommand{\smallsym}[2]{#1{\mathpalette\make@small@sym{#2}}}
\newcommand{\make@small@sym}[2]{%
  \vcenter{\hbox{$\m@th\downgrade@style#1#2$}}%
}
\newcommand{\downgrade@style}[1]{%
  \ifx#1\displaystyle\scriptstyle\else
    \ifx#1\textstyle\scriptstyle\else
      \scriptscriptstyle
  \fi\fi
}
\makeatother

\author{
  Thomas F Burns\\
  Cornell University
}

\begin{document}

\maketitle

\begin{abstract}
Southeast Asia is a geopolitically and socio-economically significant region with unique challenges and opportunities. Intensifying progress in generative AI against a backdrop of existing health security threats makes applications of AI to mitigate such threats attractive but also risky if done without due caution. This paper provides a brief sketch of some of the applications of AI for health security and the regional policy and governance landscape. I focus on policy and governance activities of the Association of Southeast Asian Nations (ASEAN), an international body whose member states represent 691 million people. I conclude by identifying sustainability as an area of opportunity for policymakers and recommend priority areas for generative AI researchers to make the most impact with their work.
\end{abstract}

\section{Introduction}

Health security has undergone a rapid integration of artificial intelligence (AI) technologies, courtesy not only of the rapid advances in AI research and deployment \citep{nestor_maslej_e97e0966}, but also the significant global response to the COVID-19 pandemic \citep{the_lancet_d3acd361, nina_schwalbe_5ed47c30, bashar_haruna_gulumbe_91d96bfa}. Southeast Asia, home to diverse nations with varying levels of economic and technological development, faces a unique set of challenges in leveraging AI for enhanced disease surveillance, treatment, and prevention, and in strengthening general healthcare infrastructure. Their success in doing so is of global interest given their unique geopolitical position. Politically wedged between the interests of the United States and China, in recent years Southeast Asian states have carefully charted a middle course, seeking to preserve their autonomy and maximise local interests amidst great power competition \citep{see_seng_tan_c9876e46, cheng_chwee_kuik_a88c3c67, nguyen_van_hiep_assoc__prof_51a8259f}. Therefore, analysing the policy and governance landscape in Southeast Asia with respect to the continued safe development and deployment of AI technologies in this domain, and promoting alignment with regional health and broader security goals, may help support regional and global discussions in these areas.

For the purposes of this paper, I will consider Southeast Asia to be the geographic area incorporating the countries of Brunei Darussalam, Cambodia, Indonesia, Lao People's Democratic Republic (PDR), Malaysia, Myanmar, the Philippines, Singapore, Thailand, Timor-Leste, and Viet Nam. In 2025, the region's population is estimated to reach 691 million people with a nominal gross domestic product of 4.4 trillion U.S. dollars \citep{asean_timor-leste_pop_gdp}. All but Timor-Leste are current members of the Association of Southeast Asian Nations (ASEAN), with a 2022 in-principle agreement to admit Timor-Leste as the eleventh ASEAN member state \citep{timor-leste_asean_membership}. The region has significant shared health security concerns \citep{marie_lamy_6389692a, riyanti_djalante_54435c94}, with ongoing efforts to strengthen regional cooperation and improve health  \citep{ endorsed_nasean_declaration_of_commitment_on_hiv_and_aids__nfast_tracking_and_sustaining_hiv_and_aids_responses_nto_end_the_aids_epidemic_by_2030_df529fa8, plan_of_action_to_implement_the_asean_leaders__declaration_on_disaster_health_management__2019_2025__c4289c67,strengthening_health_architecture_in_the_region___2023_indonesia_s_asean_chairmanship_spotlight__f788b5b4,tobacco_control_sector___fctc_secretariat_ministry_of_health__malaysia_365ec36d}.

\section{Generative AI applications in health security}

Health security can be defined as the activities, proactive and reactive, which minimise the impact of acute public health events on human populations. Such activities include but are not limited to disease surveillance, early warning systems, epidemic and pandemic preparedness and responses, strengthening general healthcare systems and infrastructure, and coordinating regional or international cooperation. The World Health Organization (WHO) has recognised the potential of AI to improve health security \citep{WHO_ethics_and_governance_of_AI, sameer_pujari_9323e3f8}, and leading global health scholars have highlighted unique challenges and opportunities for low- and middle-income contexts in the Global South \citep{brian_wahl_779af9e8, ahmed_hosny_be623f2a, nina_schwalbe_5ed47c30, hassane_alami_f73652cf, chinasa_t__okolo_44cd95fe}. AI-based technologies are thus being explored for a range of relevant applications  \citep{trevor_d__hadley_8b6f8a8b, nirupam_bajpai_20769ce1, sameer_pujari_9323e3f8}, many of which require continued international cooperation  \citep{Luengo-Oroz2020} and raise important policy and governance questions. Here I very briefly sketch some of the ways generative AI has emerged as an enabling technology in health security.

\paragraph*{Drug and vaccine development}

Advancements in generative AI have significantly enhanced the ability of researchers to develop new pharmaceutical compounds by rapidly and efficiently screening vast numbers of molecules \citep{hao_lv_8b6bbbda, sam_ad__jacobs_b25555d2}. Such screening can evaluate properties such as binding affinity, toxicity, and pharmacokinetic characteristics. These \textit{in-silico} drug discovery processes can reduce the time and cost required for the overall drug development pipeline. Furthermore, AI-assisted drug discovery has been applied to address diseases that disproportionately affect populations in low- and middle-income countries, such as trypanosomiasis and tuberculosis \citep{david_a__winkler_c17ed6d0, palle_jakobsen_cfb737b9}, where limited commercial incentives can hinder traditional, more costly pharmaceutical research and development investments.

\paragraph*{Diagnostics}

Correctly and promptly diagnosing disease is crucial for enabling more effective and rigorous public health responses, especially in low- and middle-income contexts  \citep{david_mabey_4df7f215}. In diagnostic applications, deep learning models have demonstrated high accuracy in interpreting medical data, assisting clinicians in the detection and monitoring of various diseases \citep{xiaoxuan_liu_4d0bc18f, xueyan_mei_3913ec13, val_rian_turb__a6978e94, s__kevin_zhou_e78d901f}. However, training data can be limited and in some cases researchers have proposed using generative AI technologies to bolster existing databases \citep{waheed2020covidgan,singh2021covidscreen,felipe_giuste_9f52ce38}.

\paragraph*{Treatment, education, and administration}

Large language models (LLMs) and other generative AI-powered chatbots using telemedicine platforms may help augment general healthcare capacity by providing initial triage and basic remote consultations \citep{craig_kuziemsky_4091f7b2, n__a__omoregbe_e8ea7507, mladjan_jovanovic_161ac376, urmil_bharti_c1ce5325, md_naseef_ur_rahman_chowdhury_5044c32d, chen2023impactrespondingpatientmessages, ClinicalGPT}, assisting in digital discussions for patient support groups \citep{najeeb_gambo_abdulhamid_e65b12de}, providing mental health support \citep{_brahim_sarbay_23760a2c, lorenzo_james_b8ab52db, elizabeth_c__stade_e0edf4d5,mariana_pinto_da_costa_17f91f53, nancy_s__jecker_556ba89d, w__mieleszczenko_kowszewicz_0df2e0dc, adam_s__miner_e5c77504}, assisting in healthcare worker education \citep{tiffany_kung_a1e07fe2}, and assisting healthcare workers in completing various administrative, research, or decision-making tasks \citep{yang_rui_18aff74f}.

\paragraph*{Public communication and information}

Generative AI tools can enhance public health communication and information dissemination during disease outbreaks \citep{elise_karinshak_9d5ad0a3, ziang_xiao_2c177d12}. Conversational AI systems, such as chatbots and virtual assistants, can provide the public with accurate, up-to-date information about the outbreak, including symptoms, prevention measures, and available resources \citep{manal_almalki_ec135d5b, mollie_m__mckillop_9612eefd}. Additionally, generative AI can personalise health recommendations based on individual risk factors and location, with a goal of improving compliance with and effectiveness of health advice \citep{shadi_alian_46692fd8}. Furthermore, social media monitoring tools can help identify and address misinformation and rumours, ensuring the public has access to credible information \citep{jiawei_zhou_8cc3c28e, irfhana_zakir_hussain_83eb4078}.

\paragraph*{Protection from malicious and reckless actors}

As a dual-use technology, AI may facilitate bioweapon development or deployment by nation states, violent extremists, or careless actors \citep{filippa_lentzos_bf452b49, ashima_jha_e25b1009, ronaldo_lima_a84ec359}. However, there are potential AI-powered mitigation strategies \citep{juan__cambeiro_a0a26fe9}. In particular, researchers have proposed combined generative AI and cryptographic methods \citep{dana_gretton_8d21b5b4,carsten_baum_2315046f} to limit the risk of dangerous DNA synthesis of known or potential novel pathogens.

\section{ASEAN policy and governance related to generative AI and health security}

ASEAN have developed and implemented significant health security policy and governance structures, especially in response to COVID-19. These measures have included: the establishment of a regional coordination centre for public health emergencies and emerging diseases; policies to enable timely movement of needed medical personnel and supplies; efforts to improve regional cooperation in disaster management; setting targets to achieve vaccine self-sufficiency; and programs to unite against anti-microbial resistance and endemic health challenges, while also strengthening general healthcare infrastructure and systems regionally.\footnote{For a more detailed background, see Appendix \ref{appendix:health-security}.}

The ASEAN Guide on AI Governance and Ethics (2024) \citep{asean_guide_on_ai_governance_and_ethics} is a recent landmark voluntary guidance document that outlines principles for the development and deployment of AI technologies in the region. The Guide covers key areas such as privacy protection, algorithmic bias, human control, and the social impacts of AI. In June 2024, the ASEAN Ministerial Meeting on Science, Technology and Innovation further recognised \citep{asean_ministerial_meeting_on_science__technology_and_innovation_ammsti_77db7ca8} the transformative potential of AI and projected its economic impact to increase GDP 10--18\% by 2030 within ASEAN member states. The meeting highlighted AI's applicability in areas like health, manufacturing, transportation, smart cities, and agriculture — all of which are relevant to enhancing regional health security. 

The Guide establishes a set of principles to promote the responsible development and use of AI in the region. Taken together, these guiding principles largely align with common principles found in other international guidelines \citep{anna_jobin_aa1b18d1}: transparency, justice and fairness, non-maleficence, responsibility, and privacy. Further, each of the Guide’s seven principles can be directly related to ethics and governance issues in the domain of health security:
\begin{enumerate}
\item \underline{Transparency and explainability} is needed if we want AI systems used in disease surveillance, diagnostics, or treatment planning to be auditable and their decision-making processes understood. Similarly, explainable AI in diagnostic applications builds trust among healthcare professionals and patients by clarifying the reasoning behind systems’ diagnoses and recommendations.
\item \underline{Fairness and equity} will be imperative if AI-based health interventions are to be equitably deployed across the population. This will also be an important principle for careful model evaluation to avoid potential biases and exacerbating existing health disparities, particularly in areas like resource allocation, access, and triage decision-making.
\item \underline{Security and safety} is important in the health domain given the sensitive nature of health data and the history of malicious cyberattacks targeting such data and their systems. When AI becomes more integrated into health systems, model security and safety becomes even more important for ensuring effectiveness and continuity of essential health services during times of crisis.
\item A \underline{human-centric} approach must underpin all AI applications in health security. While AI can augment and enhance decision-making, human oversight and involvement remain critical in what are deeply human systems. AI-powered diagnostic tools should also complement, not supplant \citep{thomas_burns_f88e1552}, the expertise and clinical judgement of healthcare professionals, patients, and other stakeholders.
\item Robust \underline{data privacy and governance} frameworks are also key, especially for sensitive applications like contact tracing. Transparency in data handling and user consent are precursors for building public trust in the use of these technologies.
\item \underline{Accountability and integrity} are crucial for maintaining public trust in AI-driven health security measures. Clear lines of responsibility for AI-driven or -informed decisions, along with robust integrity and audit mechanisms for both data and models, are vital to ensure ethical and responsible use.
\item \underline{Robustness and reliability} of these AI systems is necessary when deployed in critical health security situations. Rigorous testing and validation of AI models, particularly those used in high-stakes applications like early warning systems or vaccine development, will be an essential ongoing process.
\end{enumerate}

\section{Sustainability as an area of opportunity}

Considered in the health domain, the ASEAN Guide’s principles comport with some strategic principles of WHO's Global Strategy for Digital Health \citep{who_global_strategy_for_digital_health_2020_2025_b9bfc65e}, including promotion of people-centred and equitable digital health solutions, and an emphasis on data privacy. Explicit considerations of the systemic sustainability of potential AI solutions, however, remains an area of opportunity for the ASEAN Guide. The Guide also broadly aligns with WHO guidelines for the ethical use and governance of AI for health \citep{WHO_ethics_and_governance_of_AI}, like protecting patient privacy, ensuring algorithmic fairness, and maintaining human oversight and control. However, here the WHO guidance also emphasises sustainability, not only in the direct context of health systems themselves, but additionally in the environmental and energy infrastructure senses: “AI systems should be designed to minimize their ecological footprints and increase energy efficiency, so that use of AI is consistent with society’s efforts to reduce the impact of human beings on the Earth’s environment, ecosystems and climate” \citep{WHO_ethics_and_governance_of_AI} (pg. 30). Such considerations align with ASEAN health security policy \citep{APHDA_2021-2025}, which calls for preparation and response to environmental health threats, ``including the health impacts of climate change in the region.''

There are therefore two important and distinct sustainability dimensions which are identified as areas of opportunity for the ASEAN Guide's application to health security: (1) environmental; and (2) socio-cultural and economic.

\subsection{Environmental dimension}

Environmental sustainability and the ecological impacts of AI development and deployment can be considered health security concerns due to downstream effects on human health \citep{richie2022environmentally}. AI proliferation may lead to increased deforestation and habitat loss caused by construction of new datacenters or energy infrastructure in the region, increased emissions due to increased local compute for training or inference, and, ultimately, worsening climate change \citep{luers2024will}, as well as potential redirection of important water resources for cooling datacenters \citep{david_mytton_851ae5ca}. This can occur both directly via AI applied to health security, and indirectly via AI applied to other domains.

Such effects can increase the risk of zoonotic diseases crossing over from animal populations to humans and leading to future pandemics \citep{jonathan_a__patz_5052c215, covid_19__a_warning__addressing_environmental_threats_and_the_risk_of_future_pandemics_in_asia_and_the_pacific_4890f080, jamie_k_reaser_081fb62d}. On more local and immediate terms, datacenters and energy-intensive AI-devices could disrupt electrical grid reliability and increase local energy prices \citep{md__tarek_hasan_2d9991d0, Lin_2024}, causing potential energy insecurity or power outages in hospitals and health surveillance centres. There is therefore an opportunity for environmental sustainability to be added as an explicit guiding principle in generative AI applied to health security, aligning with existing ASEAN health security policy \citep{APHDA_2021-2025}.

\subsection{Socio-cultural and economic dimension}

Work by \cite{naveena_karusala_3add0730} emphasises that interventions aimed at benefitting individuals need to consider broader systemic issues to be effective. For example, healthcare workers can be overburdened by a fractured or under-resourced health system, meaning that an intervention requiring significant training or initial time investments from healthcare workers may be immediately infeasible. For generative AI researchers, this implies that designing AI-driven services should go beyond individual-level solutions, also engaging with systemic and structural challenges. This perspective could inform the development of AI systems that are more attuned to contextual and socio-cultural complexities, ultimately fostering more effective and equitable health interventions.

Unfortunately, it is common for new technologies to be introduced in healthcare settings without long-term systemic sustainability plans \citep{matthew_cripps_47695d90}, resulting in resources being wasted on (often expensive) ‘solutions’ when the resources could have been spent on more sustainable solutions with better long-term outcomes, especially in low-resource settings  \citep{abekah_nkrumah_gordon_2da61e68, muhambe_titus_mukisa_b009f935}, e.g., where a device is likely to break down or require replacement components which are not readily available or affordable \citep{malkin2007barriers}, it is likely preferable to deploy a `lower tech' solution which can be maintained self-sufficiently by local people and systems.

It is therefore commendable that the ASEAN Guide includes the principle of human-centricity, which can be interpreted as promoting the design and deployment of AI systems which prioritise human needs and values, not just in the present but also into the future. This interpretation, if shared by and emphasised in practitioner communities, and along with the principle of robustness and reliability, will assist in AI systems meeting the needs of growing populations, which may be accompanied by changing demographics and different health challenges. Taking a human-centric approach will hopefully, too, encourage comprehensive integration into existing health systems on technical, workforce, and socio-cultural levels. Likewise, the Guide’s principle of security and safety could promote the goal of building and sustaining a high-trust system, which will also require long-term sustainability planning.

Including consideration of the sustainability of the regional health--AI workforce \citep{marsan2021artificial} will also be important for successful integration of these technologies. In the Southeast Asian context, research suggests medical students in countries like Malaysia have expressed optimism about the potential of AI to enhance healthcare delivery, though gaps remain in awareness and preparedness for its implementation \citep{amy_tung_c22c5d1c}. Generally, adoption of AI in health security within low- and middle-income countries remains limited, and there are few established, local evaluation processes to guide decision-making in these contexts \citep{hassane_alami_f73652cf}. Additionally, concerns have been raised about the technical challenges of real-time data analysis, data interoperability, and the need to address other human and organisational factors impacting the uptake of AI  \citep{luis_fern_ndez_luque_4c607422}.

These factors further underscore that successful integration of advanced AI technologies will require significant investments in costly infrastructure and extensive training data, which may be lacking in some healthcare systems or in the case of novel or neglected diseases. Addressing these challenges is key to ensuring equitable access to the benefits of AI-powered technologies, particularly in resource-constrained settings where the impact could be most transformative for health security.

\section{Future research directions}

Alongside policymakers, generative AI researchers also have a role to play. Although there are many promising directions of current research which are likely to have large and positive effects on health security in Southeast Asia, I wish to emphasise the following three research priorities:

\begin{enumerate}
    \item \textit{LLMs.} There are many emerging uses of LLMs for health security, particularly for telemedicine, public communication, and triage applications. Researches should be encouraged to ensure LLM performance and deployments are linguistically, culturally, and otherwise contextually appropriate, e.g., by continuing development of low-resource language LLMs. However, more broadly, there are also accessibility gaps in many communities within ASEAN due to developing telecommunications and power infrastructure reliability. Critical LLM deployments should therefore preference low-power, on-device models.
    \item \textit{Data generation and augmentation.} Using generative AI to generate or supplement biomedical and clinical databases may improve downstream AI models trained on these data, especially for rare and neglected diseases. However, researchers should use caution in potentially biasing or over-indexing on small, non-AI generated datasets. Instead, AI researchers should seek to collaborate directly with medical experts in these efforts, and seek to carry out more rigorous, controlled experimental and theoretical work in these areas.
    \item \textit{Effective communication and logistics.} Effective health communication within communities and within health systems is crucial. Likewise, ensuring efficient logistical operations, timely supply lines, and appropriate resource allocation is needed to prevent and react to public health events. There is a potentially strong role for generative AI to assist health officials in these areas, through accurate forecasting, generation of contextually-appropriate local health advice, and real-time audio language interpretation and generation.
\end{enumerate}

\section{Conclusion}

The people of Southeast Asia face significant health security challenges, and are uniquely positioned to help facilitate global cooperation in and develop policy for AI applications to these challenges. I hope many researchers and policymakers consider strongly the potential benefits and risks associated with generative AI, and work together to affect a prosperous and healthy future for all.

\subsection*{Acknowledgements}

Thank you to Ni Made Ayu Sulastri, Siti Hasliah Salleh, Isabella Turilli, Samanvya Hooda, Miles Brundage, Angela Onikepe, Tom Mulligan, and Cullen O'Keefe for insightful feedback and stimulating conversations. Thank you to the anonymous reviewers for their constructive comments which helped improve the paper. Thank you to the organisers of the GenAI for Health Workshop @ NeurIPS 2024 for their professional service and community contributions. Thank you to Blue Dot Impact and the organisers of the AI Safety Fundamentals Governance Course for supporting my learning journey.

{
\small
\bibliographystyle{plainnat}
\bibliography{main}
}

\appendix

\section*{Appendix}

\section{ASEAN health security policy and governance}\label{appendix:health-security}

ASEAN was established on August 8, 1967, through the signing of the ASEAN Declaration (Bangkok Declaration). The Charter \citep{ASEAN_charter} provides a legal and institutional framework for the organisation, emphasising principles such as non-interference and decision-making by consensus. ASEAN's health security strategy currently focuses on strengthening regional capacity to respond to pandemics and emerging diseases \citep{asean_regional_forum_annual_security_outlook_f703a02e}. A key aspect of this strategy is the ASEAN Centre for Public Health Emergencies and Emerging Diseases (ACPHEED), which aims to improve regional preparedness and response to health crises \citep{asean_opens_secretariat_for_medical_emergencies_in_thailand_e73c80e1}. Established in 2022 and hosted in Bangkok, Thailand, ACPHEED’s work includes enhancing disease surveillance, information sharing, and coordinating pandemic response efforts. Although not directly mentioned in the latest ASEAN Regional Forum Annual Security Outlook \citep{asean_regional_forum_annual_security_outlook_f703a02e}, other sources suggest that biosecurity risks, such as those which could be generated by a malicious actor, are also a significant concern for ASEAN \citep{SEA_Biosecurity}. This comports with Article 1 of the ASEAN Charter \citep{ASEAN_charter}, which states that a purpose of ASEAN is ``[t]o preserve Southeast Asia as ... free of all ... weapons of mass destruction.''

ASEAN members have also developed multiple regional strategies, reports, and agreements which enhance health security. Mutual recognition agreements allow qualified healthcare professionals to move within the region and have their qualifications recognised \citep{ASEANMutualRecognitionArrangementonMedicalPractitioners,ASEANMutualRecognitionArrangementOnDentalPractitioners,ASEANMutualRecognitionArrangementonOnNursingServices}. The Sectoral Integration Protocol for Healthcare (2004) \citep{ASEANSectoralIntegrationProtocolforHealthcare} sets out a clear plan of action to reduce friction in the movement of health-related products and personnel, as well as enhance foreign investment opportunities in the health sector within ASEAN. Built upon this foundation, ASEAN have also developed regional strategies, reports, action plans, and other agreements \citep{APHDA_healthpubs} which seek to, among other things, eliminate rabies, combat anti-microbial resistance, respond to HIV and AIDS, end malnutrition and child stunting, strengthen primary and mental health systems, and monitor and reduce health risks stemming from the consumption and distribution of alcohol, tobacco, illicit drugs, and sub-standard or falsified medicines. In addition, three notable ASEAN Leaders’ Declarations\footnote{ASEAN Leaders’ Declarations are documents written and agreed to by states' leaders which are developed through focused, often single meetings on specific topics and include certain joint acknowledgements, recommendations, and commitments.} bear particular significance for regional health security.

The ASEAN Leaders’ Declaration on ASEAN Vaccine Security and Self-Reliance in 2019 \citep{ASEAN_leaders_vaccine-security} reiterates three ``critical elements in assuring vaccine security'': ``guaranteed procurement of vaccines through firm contracts with manufacturers''; ``secured, multi-year allocations for vaccine financing''; and ``long-term accurate forecasting of vaccine requirements.'' The Declaration further says vaccine security has been ``identified as a priority strategy for health in ASEAN'', tasks ASEAN ministers to work towards vaccine security, and calls on ``development partners, intergovernmental agencies, regional organizations, experts and other stakeholders, to support and promote regional collaboration in the development of [ASEAN vaccine security and self-reliance]''.

The ASEAN Leaders’ Declaration on Disaster Health Management (2017) \citep{ASEAN_disaster-health} includes a range of commitments to strengthen health management systems at the national and regional levels to improve health outcomes in times of emergency. The Declaration calls for closer cooperation between ASEAN members in health emergencies (including in disease surveillance), development of common regional standard operating procedures for emergency medical teams, strengthening of national health systems and infrastructure, and establishing a ``Regional Disaster Health Training Center [to design simulations and joint operations,] to increase capacities of health workers and disaster health-related personnel''.

The ASEAN Leaders’ Declaration on Antimicrobial Resistance: Combating Antimicrobial Resistance through One Health Approach (2017) \citep{ASEAN_amr} establishes a regional commitment to working across sectors and professional disciplines to combat antimicrobial resistance. Measures include improving public and professional awareness and education on the inappropriate use of antimicrobials in humans, non-human animals, aquaculture, and land crops. The Declaration also calls for ``enhancing regulatory mechanisms towards a no prescription-no antibiotic [model]'', and ``strengthening national and regional laboratory capacity, surveillance and monitoring systems for antimicrobial resistance, antimicrobial consumption and use [...], and drug residues''.

Looking forward, the ASEAN Post-2015 Health Development Agenda for 2021 to 2025 \citep{APHDA_2021-2025} has the following goals for ``Responding to All Hazards and Emerging Threats'' related to health: ``To promote resilient health systems in response to communicable diseases, emerging infectious diseases, neglected tropical diseases, and zoonotic diseases''; ``To enhance regional preparedness and response to public health emergencies and ensure effective disaster health management in the region''; and ``To prepare and respond to environmental health threats and other hazards, including the health impacts of climate change in the region.''

\end{document}